\documentclass[10pt]{iopart}

\usepackage{graphicx}   
\usepackage{mathabx}        

\usepackage[urlcolor=blue]{hyperref}      

\begin{document}

\title[Universal scaling relation for magnetic sails]{Universal
scaling relation for magnetic sails: momentum braking in the 
limit of dilute interstellar media}

\author{Claudius Gros}

\address{Institute for Theoretical Physics
Goethe University Frankfurt, Frankfurt/Main, Germany}
\ead{gros07-a-t-itp.uni-frankfurt.de}
\vspace{10pt}
\begin{indented}
\item[]May 2017
\end{indented}

\begin{abstract}
The recent progress in laser propulsion research has advanced 
substantially the prospects to realize interstellar spaceflight
within a few decades. Here we examine passive deceleration 
via momentum braking from ionized interstellar media. The 
very large area to mass relations needed as a consequence
of the low interstellar densities, of the order of 0.1 
particles per $\mathrm{cm}^{3}$, or lower, are potentially
realizable with magnetic sails generated by superconducting
coils. Integrating the equations of motion for interstellar 
protons hitting a Biot Savart loop we evaluate the
effective reflection area $A(v)$ in terms of the velocity
$v$ of the craft. We find that the numerical data is fitted
over two orders of magnitude by the scaling relation
$A(v)\ =\ 0.081A_R\log^3(I/(\beta I_c))$, where $A_R=\pi R^2$
is the bare sail area, $I$ the current and $\beta=v/c$.
The critical current $I_c$ is $1.55\cdot10^6$\,Ampere. The
resulting universal deceleration profile can be evaluated 
analytically and mission parameters optimized for a minimal
craft mass.

For the case of a sample high speed transit to Proxima Centauri 
we find that magnetic momentum braking would involve daunting 
mass requirements of the order of $10^3$\,tons. A low speed 
mission to the Trappist-1 system could be realized on the other 
side already with a 1.5\,ton spacecraft, which would be furthermore 
compatible with the specifications of currently envisioned directed 
energy launch systems. The extended cruising times of the order of 
$10^4$ years imply however that a mission to the Trappist-1 system 
would be viable only for mission concepts for which time constrains 
are not relevant.
\end{abstract}

%
\noindent{\it Keywords}: magnetic sail, Biot Savart loop, interstellar spaceflight, momentum braking
%
%
%
\ioptwocol
%

\section{Introduction}

Research in interstellar travel technologies has
seen a paradigm change over the last few years.
Early concept studies, like the Orion
\cite{dyson1969discussion} and the Daedalus
\cite{bond1978project} project, envisioned
gigantic fusion based starships \cite{long2011project}.
The perspective changed however when efforts to develop 
miniaturized satellites, i.e.\ waferSats \cite{brashears2016building}, 
were expanded to designs of spacecraft-on-the-chip probes
suitable for deep space exploration \cite{lubin2016roadmap}.
The low weight of these nanocrafts would allow to accelerate 
them at launch to a substantial fraction of the speed of light 
by ground- or space based arrays of lasers disposing of an overall
power of the order of 100\,GW \cite{lubin2016roadmap}. Other 
aspects of long-duration interstellar missions, such as
the feasibility of self-healing electronics based on 
silicon nanowire gate-all-around FET \cite{moon2016sustainable}
and the interaction of relativistic spacecrafts with the interstellar 
medium \cite{hoang2017interaction}, have also been addressed.

The ever longer list of known exoplanets \cite{schneider2011defining}
includes by now nearby potentially habitable planets 
\cite{anglada2016terrestrial,dittmann2017temperate}. 
The number of promising interstellar mission concepts
is likewise increasing, ranging presently from high speed Proxima 
Centauri flybys \cite{merali2016shooting} to slow cruising Genesis 
missions aiming to establish autonomously developing biospheres 
of unicellular organisms on transiently habitable
exoplanets \cite{gros2016developing}. These developments
warrants hence to reexamine the options available to decelerate
the interstellar probe passively on arrival.

\begin{figure}[t]
\centering
\includegraphics[width=0.90\columnwidth,angle=0]{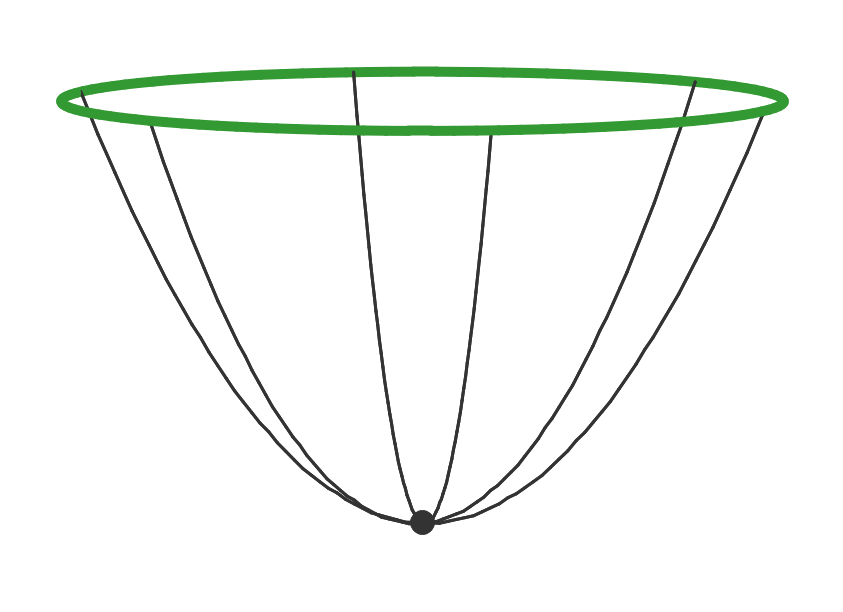}
\caption{Illustration of a magnetic sail in the form of a 
superconducting Biot Savart loop (green). Forces are transmitted 
via tethers (black lines) to the payload (filled
sphere) located in front of the sail.
        }
\label{fig:magSailCraft}
\end{figure}

Spacecrafts travel essentially unimpeded through the
voids of deep space, which is characterized by very 
low particle densities. Electromagnetic fields 
produced by the craft may however reach far enough
for the spacecraft to transfer its kinetic momentum
slowly to the interstellar medium. One speaks
of a magnetic sail \cite{zubrin1991magnetic},
as illustrated in Fig.~\ref{fig:magSailCraft}, 
for the case that a magnetic field produced by a 
superconducting Biot Savart loop does the job of 
reflecting the $H^+$ ions of the interstellar
medium.

The requirements for a magnetic sail are daunting.
Firstly, because currents of typically $10^6$ Ampere 
are needed to generate a magnetic field capable to 
reflect protons at relativistic speeds, and secondly 
due to the low particle density of typically less than 
one particle per cm$^3$ of the interstellar medium. 
Performing numerical simulations valid in the 
limit of low interstellar particle densities we
show here that a single universal scaling function 
describes the dynamics of magnetic sails to an 
astonishing accuracy. The resulting model, which
can be evaluated analytically, may hence serve
as a reference model. We also point out that 
previous estimates \cite{zubrin1991magnetic}
for the operational properties of magnetic sails
have been rather on the optimistic side.

\section{Passive interstellar deceleration}

Passive deceleration techniques of interstellar crafts
are based on transferring the momentum of the probe 
to other media. The three options for braking
are to transfer momentum to either the 
\begin{itemize}
\item interstellar medium,
\item the stellar wind of the target star or
\item to the photons of the target star.
\end{itemize}
For the last possibility a photon sail is used 
\cite{macdonald2011solar,fu2016solar}, just in the reverse 
modus operandi \cite{heller2017deceleration}. For the first 
two options momentum is transferred to charged particles, with 
the difference being that the particle density of stellar 
winds is somewhat higher than that of the interstellar space.
Passive deceleration of interstellar crafts is both 
suitable and necessary for interstellar missions for which 
the overall duration is irrelevant \cite{gros2016developing},
such as missions aiming to offer life alternative evolutionary pathways
\cite{crick1973directed,meot1979directed,mautner1997directed}.

\subsection{Interstellar medium}

The interstellar neighborhood of the sun is
characterized to a distance of about 15 lightyears
by a collection of interstellar clouds, with the 
local interstellar cloud interacting with the 
G cloud. Together they are embedded into the local 
bubble, which reaches out to roughly 300 
lightyears \cite{frisch2011interstellar}.

Interstellar clouds vary strongly in their properties.
Typical densities of ionized hydrogen are of the
order of $n_p=0.05-0.20$\,cm$^{-3}$ for the warm local 
clouds \cite{frisch2011interstellar} and about 0.005\,cm$^{-3}$ 
for the voids of the local bubble \cite{welsh2009trouble}.
Patches of isolated cold interstellar clouds with less
than 200\,AU (astronomical units) in diameter may have on 
the other hand large densities of neutral hydrogen, up to the 
order of 3000\,cm$^{-3}$ \cite{meyer2012remarkable}.

Passive deceleration of an interstellar probe based on 
transferring kinetic energy to ionized hydrogen,
i.e.\ to protons, is hence substantially more difficult 
for destinations located in the local bubble. We
investigate this problem here in the limit of
independent particles, touching shortly on
the possible formation of magnetic bow shocks in
Appendix~B.

\subsection{Impulse braking}

An interstellar craft with a total mass $m_{tot}$ and
an area $A$ perpendicular to the cruising direction 
encounters $A v n_p$ particles per second, where 
$n_p=f_{ISM}\cdot10^{6}$\,m$^{-3}$ is the density
of protons and $v=v(t)$ the velocity. The factor 
$f_{ISM}$ characterizes the interstellar medium, with 
$f_{ISM}\approx 0.1/0.005$ for the local cloud and 
respectively the local bubble. 

Impulse braking takes place when the particles
encountered are reflected. The velocity of the craft
then changes as
\begin{equation}
\frac{d}{dt} \Big(m_{tot}v\Big) \ = \
-\big(A(v) n_p v\big)\, \big(2m_p v \big)
\label{dot_v_interstellar}
\end{equation}
for the case that the incoming particles of mass $m_p$ are
fully reflected. Here we investigate how the effective reflection 
area $A=A(v)$ depends on the velocity $v$ of the craft for the 
case that the interstellar protons are reflected by the
magnetic field of a current carrying loop. Alternatively
one may consider the reflection of interstellar ions from
electric sails \cite{janhunen2004electric,perakis2016combining}.

\begin{figure}[t]
\centering
\includegraphics[width=0.95\columnwidth,angle=0]{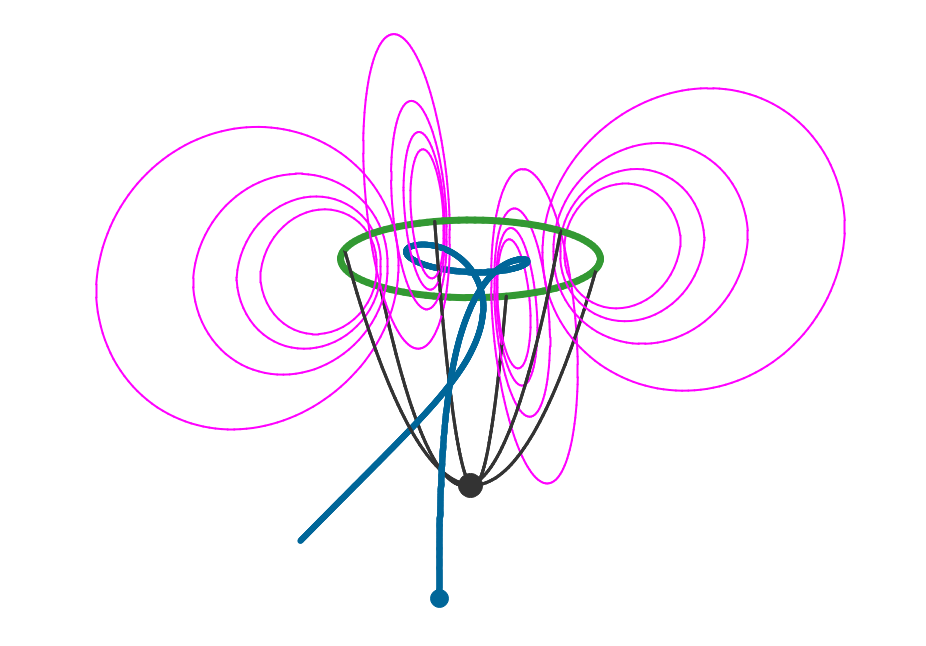}
\caption{A proton with mass $m_p$ (blue curve), starting with 
$v= 0.01\,\mathrm{c}$ with an initial offset $r=0.88\,R$ from 
the center (blue dot), is partly reflected by the magnetic field 
(pink lines) of a Biot Savart loop (green circle), carrying
in turn a current of $I=10^5$\,A. The process is scale invariant 
with respect to the radius $R$ of the loop. The normalized
scalar product $\mathbf{v}_i\cdot\mathbf{v}_f/v^2$ between the 
initial and the final velocities, $\mathbf{v}_i$ and respectively 
$\mathbf{v}_f$, is here $\mathbf{v}_i\cdot\mathbf{v}_f/v^2 = -0.61$.
The direction of flight of the spacecraft is in this perspective 
from top to down.
        }
\label{fig:magSail}
\end{figure}

\subsection{Biot Savart magnetic sail}

Charged particles may be partially reflected by
the magnetic field 
\begin{equation}
\mathbf{B}(\mathbf{x}) \ = \ \frac{\mu_0I}{4\pi} \int_L 
\frac{d\mathbf{l}'\times (\mathbf{x}-\mathbf{x}')}
{|\mathbf{x}-\mathbf{x}'|^3}
\label{B_Biot_Savart}
\end{equation}
of a current carrying loop $L$, as illustrated in 
Fig.~\ref{fig:magSail}. For a loop of radius $R$ the 
field strength $B_0=B(0)$ at the center is
\begin{equation}
B_0 \ = \ \frac{\mu_0I}{2R}~.
\label{B_Biot_Savart_center}
\end{equation}
As a reference we remark that $B_0=2\pi\cdot10^{-7}$\,Tesla 
for $R=1$\,m and a current of $I=1$\,Ampere. After testing 
for accuracy we have used for our numerical simulations
12-48 segments as subdivisions for the Biot Savart loop.
The trajectories, such as the one shown in 
Fig.~\ref{fig:magSail}, have then been evaluated by a 
straightforward numerical integration of the 
non-relativistic Lorentz force 
\begin{equation}
m_p\dot \mathbf{v}\ =\ q\,\mathbf{v}\times\mathbf{B}~,
\label{Lorentz_force}
\end{equation}
where $m_p$ and $q$ are respectively the proton mass and charge.

A particle in a homogeneous field $B$ performs
circular orbits for which the radius, the
Larmor radius $r_L$, is 
\begin{equation}
r_L \ =\  \frac{m v}{|q|B}
\ = \frac{2 m_p v}{|q|\mu_0I}\,R~,
\label{Larmor_radius}
\end{equation}
where we assumed in the second step that $B=B_0$, as given 
by (\ref{B_Biot_Savart_center}). For the case of protons
the Larmor radius (in meters) is given
by $3.13\,\beta/B[\mathrm{Tesla}]$.

A magnetic sail functions when the Larmor
radius remains within the craft diameter, viz 
when $r_L<2R$. With $\beta =v/c$ the
expression (\ref{Larmor_radius}) for the Larmor
radius then implies that
\begin{equation}
I \ >\ \frac{m_p c}{|q|\mu_0}\, \beta
\ \approx\ \beta\cdot 2.5\cdot 10^{6}\,\mathrm{Ampere}
\label{magSail_condition}
\end{equation}
needs to be fulfilled, as an order of magnitude 
estimate, if protons are to be reflected by the 
magnetic field of the loop. We note that the 
radius $R$ does not enter (\ref{magSail_condition}).

\section{Scaling of the effective reflection area}

We consider here the axial configuration, i.e.\
that the center magnetic field is aligned with 
the initial velocity $\mathbf{v}_i$ of the 
particles. Integrating the trajectories of 
protons arriving at a distance $r$ from the 
center of the Biot Savart loop we did evaluate 
the normalized scalar product between the initial 
and the final velocities, $\mathbf{v}_i$ and 
$\mathbf{v}_f$ respectively,
$$
S(r) \ =\ {\bf v}_{i}\cdot {\bf v}_{f}/v^2,
\qquad |{\bf v}_{i}|=v=|{\bf v}_{f}|~.
$$
The particle is unaffected by the magnetic sail 
for $S(r)=1$ and fully reflected for $S(r)=-1$. 

\begin{figure}[t]
\centering
\includegraphics[width=0.90\columnwidth,angle=0]{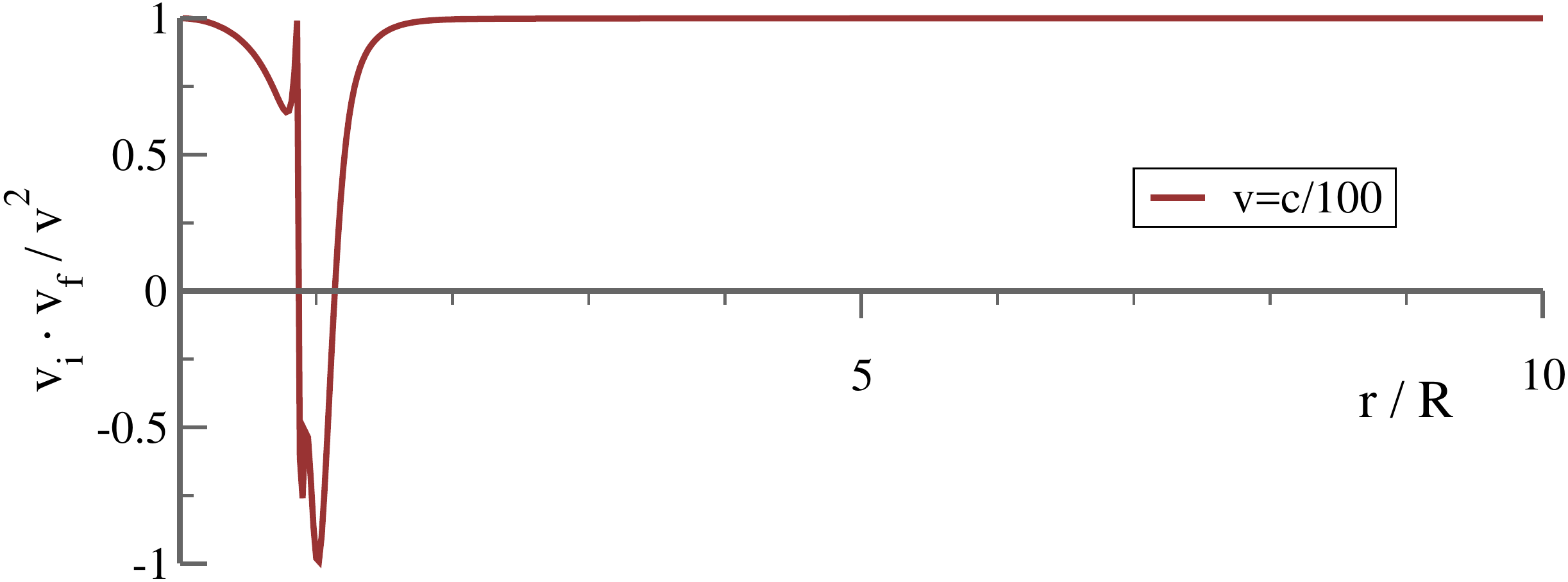}
\vspace{0.5ex}
\includegraphics[width=0.90\columnwidth,angle=0]{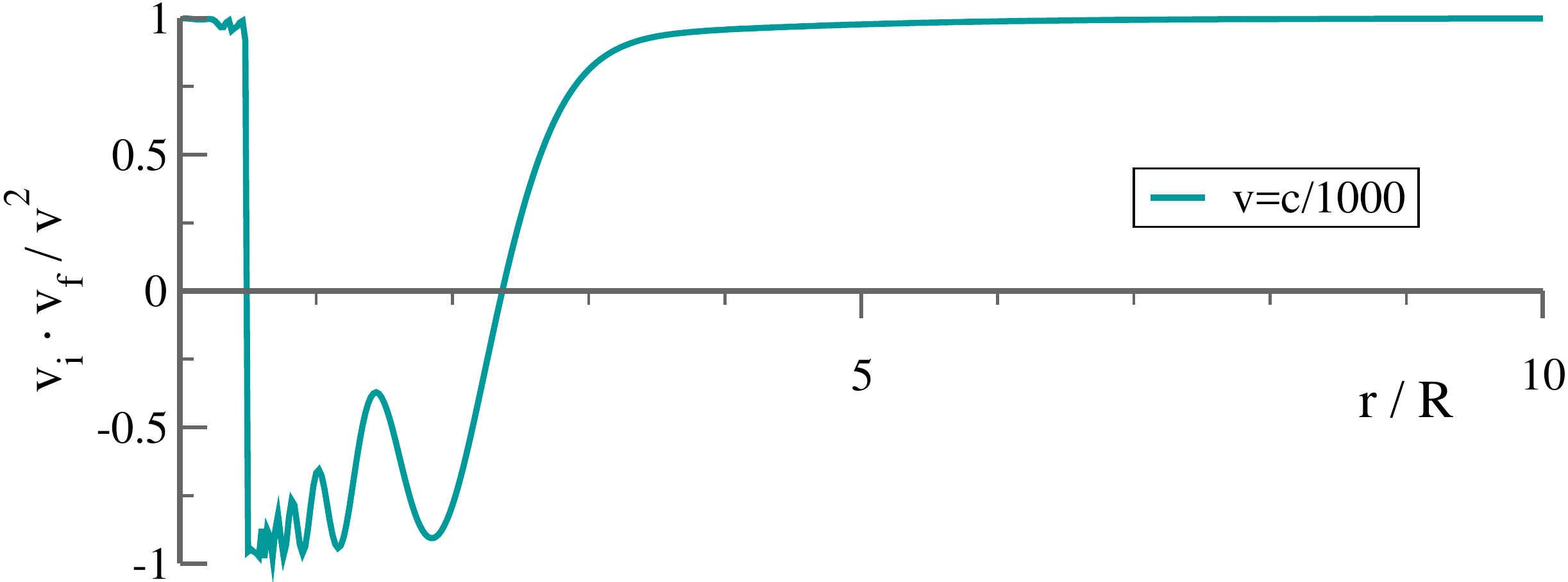}
\vspace{0.5ex}
\includegraphics[width=0.90\columnwidth,angle=0]{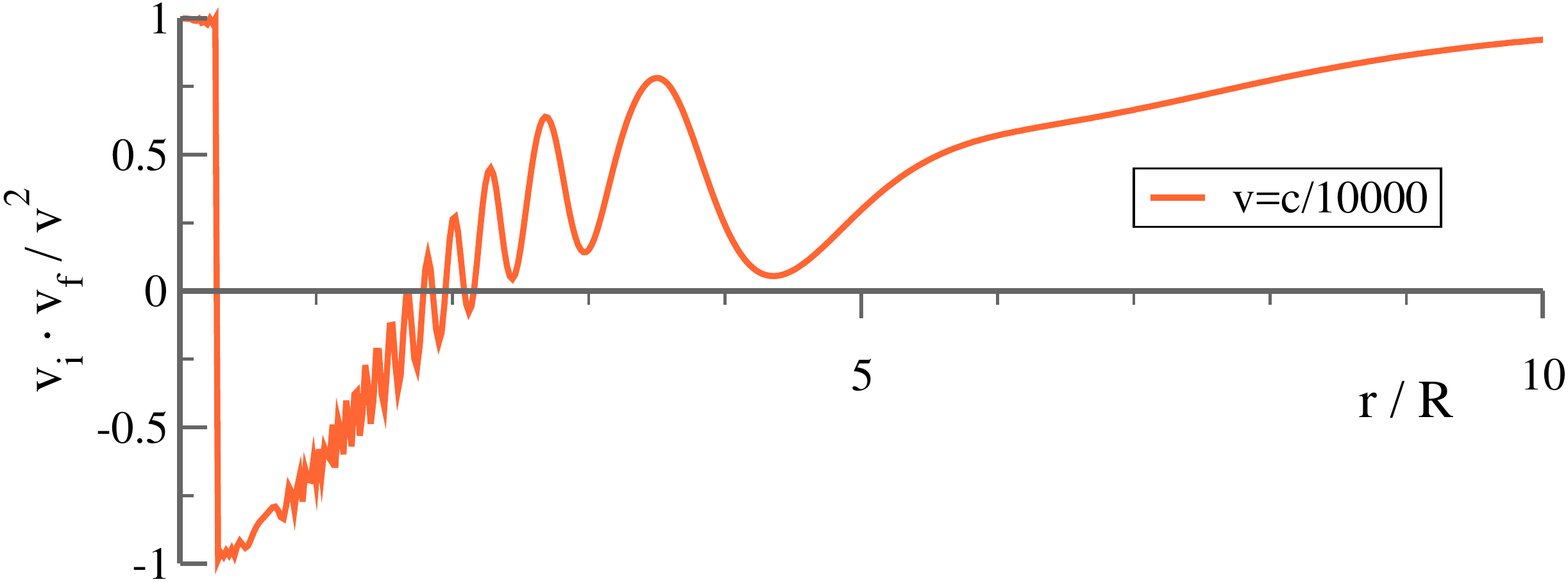}
\caption{The normalized scalar product 
$\mathbf{v}_i\cdot\mathbf{v}_f/v^2$ between
the the initial and the final velocities, $\mathbf{v}_i$ and
respectively $\mathbf{v}_f$, of a proton hitting a Biot Savart
loop with a current of $10^5$\,Ampere at a distance $r$ from the 
center.  The velocity of the proton is $c/100$, $c/1000$ and $c/10000$
from top to bottom. The magnetic field at the center is along the $z$-axis
and hence parallel to $\mathbf{v}_i$. Note the highly non-monotonic
features and that the magnetic sail becomes ineffective for
$v>cI/I_c=c/15.5=0.065c$, see (\ref{A_v_universal}). The 
normalized scalar product  $\mathbf{v}_i\cdot\mathbf{v}_f/v^2$ is 
a function of only $r/R$ and hence fully scale invariant with respect
to the loop radius $R$.
        }
\label{fig:v_i_dot_v_f}
\end{figure}

Selected results for $S(r)$ are presented in 
Fig.~\ref{fig:v_i_dot_v_f}. The alignment of
the direction of the velocity with the magnetic 
field at the center of the loop results in
a central hole, corresponding to $S(r)\to1$.
It is evident that $S(r)$ is otherwise strongly 
non-monotonic, with perfect transmission $S(r)\to1$ 
being recovered again with increasing distance $r$.
The data presented in Fig.~\ref{fig:v_i_dot_v_f}
is fully scale invariant with respect to the 
loop radius $R$, as we have confirmed numerically.

\begin{figure}[t]
\centering
\includegraphics[width=0.90\columnwidth,angle=0]{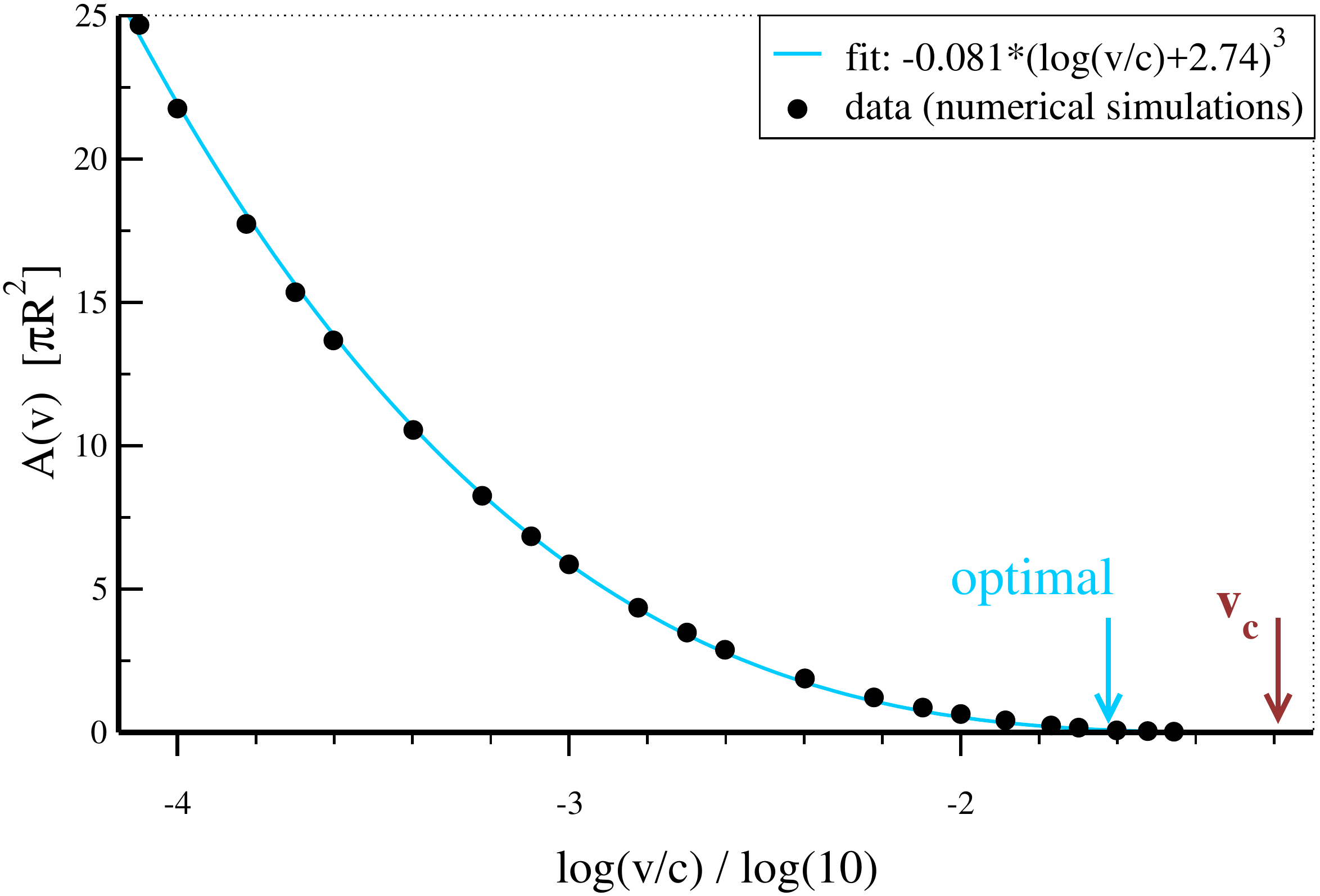}
\caption{The effective reflection area $A(v)$ of the magnetic
sail, see (\ref{A_v_int}), as obtained by simulating numerically 
the trajectories of protons hitting a Biot Savart loop (black bullets). 
The current is $I=10^5$\,Ampere. The effective area $A(v)$ is given 
in units of the bare loop area $\pi R^2$ and the velocity as
$\log(v/c)$, to base ten. The fit of the data using 
(\ref{A_v_universal}) (blue line) is excellent when optimizing 
the two parameters $A_c$ and $I_c$. The brown and blue arrows
indicate respectively the critical $v_c=c/15.5$ entering 
(\ref{A_v_universal}) and the velocity $c/42$ corresponding 
to the optimal current (\ref{I_optimal}).
}
\label{fig:A_v_logFit}
\end{figure}

The effective reflection area $A(v)$ can be 
obtained in a second step by integrating $S(r)$ 
over all distances $r$ from the center of the Biot 
Savart loop:
\begin{equation}
A(v) \ =\ 2\pi\int_0^\infty \frac{1-S(r)}{2}\,rdr~.
\label{A_v_int}
\end{equation}
The data can be fitted, as evident from the results 
presented in Fig.~\ref{fig:A_v_logFit}, to a surprising 
accuracy by the log-polynomial scaling function
\begin{equation}
A(v)\ =\ A_c\left[ \log\left( \frac{I}{\beta I_c} \right) \right]^3,
\qquad
v < (cI)/I_c~,
\label{A_v_universal}
\end{equation}
where $\beta=v/c$ is the relativistic velocity, $I$
the current through the wire and $A_c=0.081\,\pi R^2$.
The area enclosed by the Biot Savart loop is $\pi R^2$. 

The effective reflection area $A(v)$ vanishes altogether 
when $I<\beta\,I_c$\, Ampere, where the numerical result 
(\ref{A_v_universal}) for the critical current, 
$I_c=1.55\cdot10^{6}$\, Ampere, reflects the order of magnitude 
estimate (\ref{magSail_condition}). We note that we cannot 
resolve numerically whether $A(v)$ really vanishes for 
$v>cI/I_c$, or if it drops to finite but exceedingly small 
values. 

\begin{itemize}
\item The data presented in Fig.~\ref{fig:A_v_logFit}
      for $I=10^5$\,Ampere scales with $v/I$, as we have 
      verified numerically. This scaling, which has been
      included in (\ref{A_v_universal}), results from
      the linear scaling of the Lorenz force with the 
      magnetic field $B$, which increases in turn
      linearly with the current $I$. The radius $R$
      of the Biot-Savat loop only enters as a prefactor.
\item The functional form of the scaling relations 
      (\ref{A_v_universal}) cannot be traced back, to our 
      knowledge, to underlying physical arguments. It is
      however interesting to point out that a powerlaw 
      scaling \cite{zubrin1991magnetic} such as
      $A_V\sim (v/c)^\alpha$, would be inconsistent with 
      the existence of a critical current $I_c$. The 
      log-argument $\log(I/(\beta I_c)$ in (\ref{A_v_universal}) 
      captures on the other hand the fact that the effective
      reflection area $A_v$ vanishes for $v/c>I/I_c$.
\item In an equivalent study of magnetic sails in the axial 
      configuration, Andrews and Zubrin \cite{andrews1990magnetic} 
      found $A(v)\approx75$ for $R=100$\,km, $B_0=10^{-5}$\,Tesla 
      and $v=750$\,km/s, corresponding to $I=1.6\cdot10^6$\,Ampere 
      and $v=c/400$ respectively. Using identical parameters we 
      find with $A(v)=18$, a substantially more conservative estimate.
      We also note that Andrews and Zubrin \cite{andrews1990magnetic} 
      proposed scaling relations which are inconsistent with the
      here presented results. The source for these discrepancies are 
      unclear.
\end{itemize}
We note that the  effective reflection area $A(v)$ coincides 
with the loop area when $A(v)=\pi R^2$, viz when
$I/\beta = 15.6\cdot10^6\,\mathrm{Ampere}$~.

\subsection{Universal deceleration profile}

Substituting $A(v)$ from (\ref{A_v_universal})
into the equation of motion (\ref{dot_v_interstellar})
for the craft we find
\begin{equation}
\frac{\dot v/v}{\log^3(v I_c/(cI))} \ =\ 
\frac{2m_p n_p A_c}{m_{tot}}\,v ~,
\label{dot_v_A_v_universal}
\end{equation}
which can be solved for $v=v(x)$ as
\begin{eqnarray}
\nonumber
\frac{1}{\log^2(v I_c/(cI))} & =& \frac{1}{\log^2(v_0 I_c/(cI))} 
\\ & +& \frac{4m_p n_p A_c}{m_{tot}}\,(x_0-x) ~,
\label{dot_v_A_v_universal_integrated}
\end{eqnarray}
where $v_0=v(x_0)$ is the initial velocity. In the
following we set $x_0=0$. The speed drops to zero
at a distance $x_{max}$, which can be expressed as
\begin{equation}
x_{max} \ =\ \frac{1/0.081}{\log^2(v_0 I_c/(cI))} 
\,\frac{1}{4m_pn_p}\,\frac{m_{tot}}{\pi R^2}~.
\label{stopping_distance}
\end{equation}
Here we have used that $A_c=0.081\pi R^2$. The 
stopping distance $x_{max}$ may be used to
write the determining equation (\ref{dot_v_A_v_universal_integrated}) 
for the velocity as
\begin{equation}
v =  v_0\,e^a \exp\left(\frac{-a}{\sqrt{1-x/x_{max}}}\right),
\quad x\in[0,x_{max}]\,,
\label{speed_profile}
\end{equation}
where $a = \log(cI/(v_0I_c))>0$. Reversing
(\ref{speed_profile}) one find
\begin{equation}
\frac{x}{x_{max}} \ =\ 1 -\frac{a^2}{(\log(v/v_0)-a)^2}~.
\label{distance_profile}
\end{equation}
The speed profile (\ref{speed_profile}) determines 
via $dt=dx/v$ also the time 
\begin{equation}
t \ =\ \frac{e^{-a}}{v_0}\int_0^x dx'
\exp\left(a/\sqrt{1-x'/x_{max}}\right)
\label{time_profile}
\end{equation}
needed to cruise a given distance $x$. Note that $a>0$ 
and that (\ref{time_profile}) hence diverges for $x\to x_{max}$.

\subsection{Optimal current\label{sec_optimal_current}}

A straightforward route to optimize mission 
profiles is to minimize the stopping distance 
(\ref{stopping_distance}) for a given total 
mass $m_{tot}$. We assume here that $m_{tot}$
is dominated by the mass of the magnetic sail, which 
is in turn proportional to $IR$. We need hence
to minimize
\begin{equation}
\frac{z^2}{\log^2(z)}, \qquad
z = \frac{I}{\beta_0 I_c}
\label{stopping_distance_optimization}
\end{equation}
as a function of $z$. The minimum of 
(\ref{stopping_distance_optimization}), 
$\log(z)=1$, $z=e\approx 2.71$, 
determines then the current 
\begin{equation}
I_{opt} \ =\ \beta_0\cdot4.2\cdot10^6\,\mathrm{Ampere}
\label{I_optimal}
\end{equation}
minimizing the stopping distance $x_{max}$ for a given 
total mass $m_{tot}$. The resulting effective reflection
area (\ref{A_v_universal}) is
\begin{equation}
A(v) \ =\ A_c\left[1+\log(v_0/v)\right]^3,
\qquad I= I_{opt}~,
\label{A_v_opitimal}
\end{equation}
where we have used that $\log(ev_0/v)=1+\log(v_0/v)$.
Deceleration starts therefore with a very small effective 
area, viz with $A(v_0)/(\pi R^2)=0.081$.

Using the optimality condition (\ref{I_optimal}), that
is 
\begin{equation}
a\ =\ \log(z)\big|_{z=e}\ =\ 1~,
\label{a_opitimal}
\end{equation}
one may evaluate the deceleration 
profile for a given deceleration ratio $v/v_0$.
\begin{itemize}
\item From (\ref{distance_profile}) we find that
      $v/v_0=1/10$ is reached at $x=0.91\,x_{max}$. For 
      the time the craft needs to decelerate one then 
      obtains, when integrating (\ref{time_profile}) 
      numerically, $t=1.84\,x_{max}/v_0$.
\item After a cruising time of $5.2\,x_{max}/v_0$ the 
      craft arrives later on at $x=0.98\,x_{max}$, at 
      which point its speed is reduced to $v_0/300$.
\end{itemize}
Note that $x_{max}/v_0$ is the time the craft
would need to cover the stopping distance $x_{max}$
at its nominal speed $v_0$.

\section{Sample mission profiles}

In the following we use the optimal current
(\ref{I_optimal}), for which 
$a=\log(I/(\beta_0I_c)) = 1$. The stopping 
distance (\ref{stopping_distance}) may be written
consequently as
\begin{equation}
x_{max} \ =\ \frac{3.1}{m_pn_p}\,\frac{m_{tot}}{\pi R^2}~.
\label{x_max_optimal}
\end{equation}
In the next step we determine suitable $x_{max}$ 
from a time balancing principle.

\subsection{Balancing cruising and deceleration}

We consider here mission profiles for which the 
time $t_{cruise}$ and $t_{stop}$ the craft spends 
respectively for cruising and for deceleration 
are the same order of magnitude. For a 2:1 ratio
we have that $t_{cruise}=2\,t_{stop}$. The overall 
time to arrive to the destination is $t_{cruise}+t_{stop}$.

For concreteness we consider two 
distinct sample missions.
\begin{itemize}
\item[PC] A high speed mission to Proxima Centauri
          with $\beta_0=1/10$. The distance $D$ to 
          destination is $4.2$\,lyr.
\item[T1] A slow speed mission to Trappist-1
          characterized by $v_0=c/300=1000$\,km/s 
          and $D=39.5$\,lyr \cite{gillon2017seven}.
\end{itemize}
We demand in both cases that the insertion velocity 
is $c/3000=100$\,km/s. The final velocity reduction 
to planetary velocities is in this scenario assumed 
to be performed by braking from stellar wind protons 
or photons. We need hence velocity reduction factors 
$v/v_0$ of 1/300 for the PC mission and of 1/10
for T1. We use further that the respective deceleration 
distances $x$ are close to the stopping distance, i.e.\
that $x\approx x_{max}$, as discussed in 
Sect.~\ref{sec_optimal_current}. The balancing of
the cruising and the deceleration time then determines 
$x_{max}$ via
$$
t_{cruise}=2\,t_{stop},\qquad
\frac{D-x_{max}}{v_0} = 2\,\tau(v/v_0)\frac{x_{max}}{v_0}~,
$$
where $\tau(1/10)=1.84$ and $\tau(1/300)=5.2$.
The stopping distance is then 
\begin{equation}
x_{max}=\frac{D}{1+2\tau(v/v_0)} \ =\ 
\left\{
\begin{array}{rl}
0.37\,\mathrm{lyr} & \mathrm{PC} \\
8.44\,\mathrm{lyr} & \mathrm{T1} 
\end{array}
\right.
\label{x_max_PC_Ta}
\end{equation}
for the Proxima Centauri and respectively
for the Trappist-1 mission. The total 
traveling times are $58$\,year for PC
and 12000\,years for T1.

\subsection{Mass requirements}

The mass $m_{MS}$ of the magnetic sail is
\begin{equation}
m_{MS} = 2\pi R \,\frac{I\rho_I}{10^6}~,
\label{rho_I}
\end{equation}
where $\rho_I$ is the mass density per meter 
of a wire supporting $10^6$\,Ampere. We will
use in the following $\rho_I=0.18$\,kg/m,
which is, as discussed in Appendix~A,
a value one can reach with state of the art
high temperature superconducting tapes.

For the total mass $m_{tot}$ of the spacecraft, 
including tethers and payload, we assume here
that it is roughly the double of the mass $m_{MS}$
of the magnetic sail. Using (\ref{I_optimal}), 
(\ref{x_max_optimal}) and (\ref{rho_I}) 
 we then find
\begin{equation}
R = \ 0.6\cdot10^3\,\frac{\beta_0}{f_{ISM}\,x_{max}[\mathrm{lyr}]}
\label{R_final}
\end{equation}
for the radius $R$ of the magnetic sail (in km).
For the PC and the T1 mission respectively
we have $f_{ISM}=0.1/0.005$ and hence
\begin{equation}
R \ =\ 
\left\{
\begin{array}{rl}
1600\,\mathrm{km} & \mathrm{PC} \\
47\,\mathrm{km} & \mathrm{T1} 
\end{array}
\right.
\label{radius_PC_TA}
\end{equation}
and respectively $m_{tot}=1500$\,tons for PC and 
$m_{tot}=1.5$\,tons for T1. To decelerate fast 
interstellar spacecrafts is exceedingly demanding.
Above results show however also that it is possible 
to transfer the kinetic energy of slow cruising 
probes to the interstellar medium even when it is 
rarefied as within the local bubble. The craft 
parameters, $v_0=c/300$ and $m_{tot}=1.5$\,tons, would 
fit furthermore the specifications of currently envisioned
directed energy launch systems \cite{brashears2015directed}.

Slow cruising interstellar probes would need of the order of 
$10^4$ years to return data, which is clearly not a viable 
option for a science mission. The possibility to explicitly
optimize the performance of the sail, i.e.\ the effective 
reflection area (\ref{A_v_universal}), has allowed us on the 
other hand to show that interstellar missions not limited by 
time constraints \cite{gros2016developing} are feasible.
Passive deceleration using magnetic sails is in this
case possible with low-mass crafts.

\subsection{Constant time trajectories}

Instead of balancing the times needed to cruise and
to decelerate one may also consider mission profiles 
characterized by a given total travelling time 
$T_D=t_{cruise}+t_{stop}$. The stopping distance 
is then given by
\begin{equation}
x_{max} \ =\ \frac{v_0T_D-D}{\tau(v/v_0)-1}~.
\label{x_max_T_D}
\end{equation}
For a $T_D=100$\,years transit to Proxima Centauri
we $x_{max}=1.4$\,lyr, $R=430$\,km and $m_{tot}=400$\,tons.
The improvement with respect to (\ref{radius_PC_TA})
is about a factor four.

\section{Maneuvering in stellar winds\label{stellarWind}}

The velocity $v_{SW}$ of the solar wind is
$v_{SW}\sim(400-750$)\,km/s for equatorial winds in the 
ecliptic and polar winds respectively \cite{mccomas2003three}.
Both the velocity and the density $n_{SW}$
of the solar wind depend strongly on the state
of the sun in terms of the sunspot cycle. As a 
function of the distance $r$ from the star the
particle density $n_{SW}=n_{SW}(r)$ of a stellar
wind falls off like
\begin{equation}
n_{SW} = n_{AU} \frac{r_{AU}^2}{r^2},
\qquad n_{AU}\approx (1-10)\left[\frac{1}{\mathrm{cm}^3}\right],
\label{eq:solarWind}
\end{equation}
where AU stands for astronomical unit.

Magnetic braking using the stellar wind is only possible
when the luminosity of the central star is dim enough
that it does not heat up the superconducting material.
This is the case in the outskirts of M dwarfs systems, 
for which modeling efforts \cite{johnstone2015stellar} 
indicate that both the wind velocity and the proton 
density $n_{AU}$ at 1 AU distance increase with a decreasing 
mass of the star. For conservative estimates of the wind
properties of M dwarfs one may hence use solar system 
parameters.

The temperature $T$ of a passively cooling craft is given
by the per surface area balance 
\begin{equation}
 2\sigma T^4 \ =\ 
\frac{L}{L_{\Sun}}
\left(\frac{R_{AU}}{r}\right)^2
W_{\Sun}
\label{radiation_balance}
\end{equation}
between twice the emitted black body and the absorbed
stellar radiation. $W_{\Sun}=1400\,\mathrm{W}/\mathrm{m}^2$ 
is here the solar constant, $L$ and $L_{\Sun}$ the luminosity of
the target star and the sun respectively and 
$\sigma = 5.67\cdot10^{-8}\,\mathrm{W}\mathrm{m}^{-2}\mathrm{K}^{-4}$
the Stefan Boltzmann constant. For a conservative estimate we
have assumed that the reflectivity vanishes.

An operating temperature of $T=30$\,K is reached for a
craft in the Trappist-1 system, for which $L= 0.00052 L_{\Sun}$
\cite{gillon2017seven}, when $R>2.8R_{AU}$. A
superconducting material with a critical temperature 
of $T=90$\,K would allow to operate the magnetic 
sail even further in, albeit only when reducing 
correspondingly the current $I$, down to $R=0.3R_{AU}$.

\subsection{Stellar wind pressure vs.\ gravitational pull}

Both the gravitational attraction
\begin{equation}
F_{G} = \frac{G m_{tot}m_{S}}{r^2}
\label{F_G}
\end{equation}
and the force
\begin{equation}
F_{SW} = 2A(v+v_{SW}) m_p (v+v_{SW})^2 n_{AU} \frac{r_{AU}^2}{r^2}
\label{F_SW}
\end{equation}
the craft experience from reflecting the protons
of the stellar wind decay with $1/r^2$. Note that the
stellar wind velocity $v_{SW}$ adds in (\ref{F_SW})
to the relative velocity between the craft and the 
protons. For an estimate of magnitude of the $v=0$ ratio
\begin{eqnarray*}
\frac{F_{G}}{F_{SW}} & =& 
\frac{1}{2m_pv_{SW}^2n_{AU}}\,
\frac{Gm_S}{r^2_{AU}}\,
\frac{m_{tot}}{A(v_{SW})}
\\ &=&
\frac{1}{6.2v_{SW}^2}\, 
\frac{n_p}{n_{AU}}\, 
\frac{Gm_S}{r^2_{AU}}\,
x_{max}
\frac{\pi R^2}{A(v_{SW})}
\end{eqnarray*}
between $F_G$ and $F_{SW}$ we use (\ref{x_max_optimal}) 
and the solar system
values $n_{AU}=5\,\mathrm{cm}^{-3}$ and $v_{SW}=500$\,km/s
\cite{johnstone2015stellar}. One obtains
\begin{equation}
\frac{F_{G}}{F_{SW}} \ =\ 7\, f_{ISM}\,
\frac{m_{S}}{m_{\Sun}}\, \frac{\pi R^2}{A(v_{SW})}\,x_{max}[\mathrm{lyr}]~,
\label{ratio_forces}
\end{equation}
where the mass $m_{S}$ of the central star has been 
measured in terms of the solar mass $m_{\Sun}$. The 
condition (\ref{ratio_forces}) for the force balance 
is fulfilled for most  magnetic sails tailored for 
interstellar braking. For the case of the T1 mission we
obtain, when using $v_0/v_{SW}=2$,
$m_{S}/m_{\Sun}\sim 0.1$ and (\ref{A_v_opitimal}),
that $F_{G}\sim 0.08 F_{SW}$. The craft may hence 
maneuver freely within the outskirts of the 
Trappist-1 system defined by the radiation
balance (\ref{radiation_balance}).

\section{Conclusions}

Simulating the motion of individual particles
in the magnetic field generated by a current
carrying loop we found that the effective reflection
area can be fitted to a surprising accuracy by
a single scaling function. Our simulations
accurately describe magnetic sails in the limit
of low interstellar particle densities. The 
universal deceleration profile resulting from the
log-polynomial scaling function found for the
effective reflection area may hence be considered 
as a reference model for magnetic momentum braking.

Our results are valid for an idealized magnetic sail
in the axial configuration. The study of additional effects 
like the influence of the thermal motion of the interstellar 
protons on the performance of the magnetic sail is also 
possible  Preliminary simulations show in this respect that 
the effective reflection area remains unaffected at high 
velocities, shrinking however when thermal and craft 
velocities start to become comparable.

Magnetic sails fail to operate when the magnetic
field generated by the current through the loop
is too weak to transfer momentum to the interstellar 
protons. The technical feasibility of magnetic sails is 
hence dependent in first place on the availability of 
materials able to support elevated critical currents.
Using the properties of state of the art $2^{\mathrm{nd}}$ 
generation high temperature superconducting tapes and
otherwise conservative estimates for the mission parameters 
we find that magnetic sails need to be massive, in the
range of $10^3$\,tons, in order to be able to decelerate
high speed interstellar crafts. 

Optimizing the requirements for a trajectory to the 
Trappist-1 system we found importantly that 
magnetic braking is possible for low-speed crafts
even when the target star is located within the local bubble,
that is when the particle density of the interstellar
medium is as low as 0.005\,cm$^{-3}$. In this case
a 1.5\,ton craft could do the job. The launch requirements
of such a craft would be furthermore compatible with 
the specifications \cite{brashears2015directed} of the 
directed energy launch system envisioned by the 
Breakthrough Starshot project \cite{merali2016shooting} 
for Centauri flybys, which could hence see dual use.
The extended cruising time of the order $10^4$ years 
implies however that slow cruising trajectories are only 
an option for mission, such as for life-carrying Genesis 
crafts \cite{gros2016developing}, not expected to yield 
near-term results in terms of a tangible scientific 
return.

\section*{Acknowledgments}

The author thanks Xavier Obradors for extended discussions
regarding the developmental status of superconducting
wires and Adam Crow and Roser Valent\'i\ for comments.

\section*{Appendix A: Materials for superconducting wires}

State of the art coated conductors for power
applications are produced in the form of tapes,
where a superconducting YBCO layer is sandwiched 
between a metallic substrate and a top cover 
\cite{obradors2014coated}. Critical currents 
of the order of $(3-7)\cdot10^6$\,Ampere/cm$^2$ are
routinely achieved \cite{leroux2015rapid}. 
These  $2^{\mathrm{nd}}$ generation high temperature
superconducting tapes would be suitable
for magnetic sails.

\begin{itemize}
\item The overall mass requirements for superconducting 
      tapes depends on the exact specification, like the material
      and the thickness used for the metallic substrate.
      Taking here the specific mass $8.9\,\mathrm{g}/\mathrm{cm}^3$ 
      of copper as a reference, we find that a wire capable
      to support a current of $10^6$\,Ampere would weight
      about $(8.9/5)$\,g/cm length, which implies that
      $\rho_I=0.18$\,kg/m. Compare (\ref{rho_I}).
\item Envisioned operating temperatures of 30 Kelvin or
      lower \cite{leroux2015rapid} fit well with 
      deep-space environments in which the superconducting
      wire would cool down passively via black body radiation.
\item The metallic substrate of the superconducting sandwich
      is generically designed to take up a substantial amount
      of magnetomechnical stress. Additional supporting structures 
      besides the tethers are hence not necessary.
\end{itemize}
The quality of the superconducting tapes is the key
parameter determining overall mission requirements.
The hard cutoff (\ref{magSail_condition}) determines 
directly the minimal current at which the sail
stops working, with a reasonable working regime
for the current being given by (\ref{I_optimal}). 
For a given type of superconducting tape the mass per 
length necessary to support the field generating current 
is therefore dependent solely on the initial velocity $v_0$.

\section*{Appendix B: Magnetic bow shocks}

The field of the Biot Savart loop on the axis, taken
here as the z-axis, is
\begin{equation}
B(z) \ =\ \frac{\mu_0I}{2R}\,\frac{R^3}{\left(z^2+R^2\right)^{3/2}}
\ \approx\ B_0\,\frac{R^3}{z^3}~,
\label{B_Biot_Savart_z_axis}
\end{equation}
where we have taken the large $z$ limit in the second 
step. A bow shock occurs when a current carrying layer 
of charged particles forms upstream from the craft. The
magnetic field generated by this current then compensates
the magnetic field on the outside, doubling it
however on the inner side. The position of the
bow shock it determined in order of magnitude
by the equilibrium between magnetic and kinetic 
pressures \cite{zubrin1991magnetic}, respectively
when the corresponding energy densities match:
\begin{equation}
n_pm_p \frac{v^2}{2} \ =\ \frac{(2B)^2}{2\mu_0}
\ \approx\ \frac{2B_0^2}{\mu_0}
\,\frac{R^6}{z^6}~.
\label{bow_shock_z}
\end{equation}
This is clearly only a rough estimate. For the
earth the surface field $B_0\approx 20\cdot10^6$\,Tesla
interacts further out with the solar wind, for which
$n_p\approx5\,\mathrm{cm}^{-3}$ and $v\approx 450$\,km/s.
One finds that $z/R\approx26$ whereas the the bow
shock of the earth forms at around 15 radii.

\subsection{Larmor radius at shock position}

A bow shock can establish itself only if the magnetic 
field $B(z)$ is strong enough to deflect the protons
within the dimensions of shock \cite{ashida2013two}, 
i.e.\ when $r_L<z$. Using (\ref{Larmor_radius}) for the
Larmor radius $r_L=r_L(z)$ we hence need that
\begin{equation}
\frac{m_p v}{|q|2 B(z)} < z,
\qquad
\frac{z^2}{R^2} < \frac{\mu_0qI}{m_pv}
 = \frac{\mu_0qI_{wr}}{m_pc}\frac{cI}{vI_{wr}}~,
\label{bow_shock_Larmor_condition}
\end{equation}
where we have taken $2B(z)$ as the inner magnetic 
field strength and the working-regime current 
$I_{wr}$ from (\ref{I_optimal}).
For the earth the Larmor radius of solar wind protons 
at the shock location $z=15\,R$ is with $\sim24$\,km
much lower than the extension of the bow shock.
Plugging in the numbers we find that the Larmor 
radius $r_L$ will be smaller than the extend $z$ 
of the bow shock if
\begin{equation}
z \ <\ 2.5\,R\left(\frac{cI}{vI_{wr}}\right)^{1/2}
\label{bow_shock_Larmor_condition_numeric}
\end{equation}
holds. Note that $cI/(vI_{wr})=1$ at the boundary 
of the working regime.  

\subsection{Bow shock condition}

We rewrite the expression (\ref{bow_shock_z}) 
for the position of the bow shock with the help 
of (\ref{B_Biot_Savart_z_axis}) as
\begin{equation}
\frac{z^6}{R^6}  = \frac{\mu_0I^2}{R^2n_pm_pv^2}
=
\frac{\mu_0I_{wr}^2}{R_0^2n_pm_pc^2}\,
\frac{R_0^2}{R^2}
\left(\frac{cI}{vI_{wr}}\right)^2~,
\label{bow_shock_z_reformulated}
\end{equation}
where we have used $c$, $I_{wr}$ and $R_0=100$\,km
as scales. With these number we obtain
\begin{equation}
\frac{z}{R}\ = \
\frac{2.4}{\left(f_{ISM}\right)^{1/6}}
\left(\frac{R_0}{R}\right)^{1/3}
\left(\frac{cI}{vI_{wr}}\right)^{1/3}
\label{bow_shock_z_numerical}
\end{equation}
for the distance $z$ of the bow shock from the
craft.

Comparing with (\ref{bow_shock_Larmor_condition_numeric})
we observe that the two conditions (\ref{bow_shock_z_numerical})
and (\ref{bow_shock_Larmor_condition_numeric}) are only
marginally compatible. It is hence not clear to which extend
a bow shock forms. Detailed investigations of the magnetosphere 
of the craft need therefore numerical methods which go beyond 
magneto hydrodynamics \cite{ashida2013two,funaki2009research}.

A precondition for a bow shock to form is that the interstellar 
protons are deflected in the first case. Zubrin and Andrews 
proposed however, as based on a plasma fluid model for the bow 
shock, that a spacecraft characterized by 
$I=159\cdot10^3$\,Ampere and $v=c/10$  could be decelerated 
effectively by magnetic sails \cite{zubrin1991magnetic}.
Our analysis, see (\ref{A_v_universal}), indicates in contrast
that a craft with this kind of mission parameters would
be right on the point where protons are not reflected at 
all, viz on the point where the effective reflection area 
$A(v\to v_c)$ vanishes.

\section*{References}


\end{document}